# Experimental Demonstration of Probabilistic Spin Logic by Magnetic Tunnel Junctions

Yang Lv, Robert P. Bloom, and Jian-Ping Wang
Department of Electrical and Computer Engineering, University of Minnesota,
Minneapolis, MN, USA, email: lvxxx057@umn.edu, jpwang@umn.edu

*Abstract—* The recently proposed probabilistic spin logic presents promising solutions to novel computing applications. Multiple cases of implementations, including invertible logic gate, have been studied numerically by simulations. Here we report an experimental demonstration of a magnetic tunnel junction-based hardware implementation of probabilistic spin logic.

## I. INTRODUCTION

Magnetic tunnel junctions (MTJs) have been explored for reconfigurable and programmable computing [1] and true in-memory computation [2]. Recently, probabilistic spin logic (PSL) involving MTJ has been proposed as a promising solution in novel computing scenarios [3]. An MTJ-based 'p-bit' is its basic element whose state fluctuates randomly but the average of its state responds to an input. Certain functionalities, such as an invertible logic gate, can be implemented by properly coupling multiple p-bits together.

Realization of the MTJ-based PSL demands a means to individually tune each MTJ to compensate device-device variations. In our previous work, we demonstrated effective control of a thermally stable MTJ for random signal generation [4]. As shown in fig. 1, an MTJ is biased by a constant voltage source and is under a static bias magnetic field. The field activates switching from anti-parallel (AP) to parallel (P) state. When in the P state, the MTJ has lower resistance and draws more current. More current generates more spin transfer torque (STT), which activates switching from P to AP. It is shown that the average dwell time of an MTJ in P and AP state can be tuned individually by combinations of bias voltage and field. This scheme of operation allows us to experimentally demonstrate a p-circuit consisting of three MTJs in this work, despite their property variations. Additionally, this operation scheme of MTJ is also expected to be applicable for thermally unstable MTJs.

## II. DEMONSTRATION OF P-BIT BUILDING BLOCK

### A. p-Block Hardware Design and Implementation

Fig. 2(a) shows our MTJ-based hardware implementation of a p-bit building block, which we will refer as 'p-block' in short. The input signal is attenuated and shifted by resistors $R_1$, $R_2$ and $R_3$, and then is fed to the comparator as a threshold signal. The constant D.C. voltage source $V_{block}$ drives the MTJ via a small resistor $R_{sense}$ in series to randomly oscillate. The added small resistor $R_{sense}$ allows the MTJ's oscillation to be converted into a random voltage signal. This signal is then low-pass filtered by $R_{filt}$ and $C_{filt}$ before entering the comparator. The filtering allows for optimization of p-block transfer curve. Fig. 2(b) shows a photograph of the actual printed circuit board with an MTJ die mounted and wire-bonded demonstrating the p-block and further the p-circuit of invertible AND gate. The static bias field is provided by an array of permanent magnets.

Fig. 3 summarizes basic information and properties of MTJs used in this work. They are single barrier in-plane CoFeB/MgO MTJs. They are thermally stable and exhibit approximately 10 mT coercivity. They can be switched by current via STT effect.

### B. p-Block Functionality Highlight

As shown in Fig. 4, the output of the p-block is seemingly random digital signal (blue). However, its mean or time average (red) is a function of input. A positive input will pull the output average to deviate from 0.5 and decrease towards 0, while a negative input will push output average towards 1.

### C. Effect of Filtering on p-Block Response

Fig. 5 shows the waveform of $V_{mtj}$, which is the voltage across $R_{sense}$ reflecting MTJ's random oscillation. Without low-pass filtering (black, $t_C$=0 μs), sharp transitions in the MTJ's waveform negatively impact the transfer curve of the p-block due to high distributions of $V_{mtj}$ at two power rails and low distributions in between. With increasing low-pass filtering (yellow and then red), waveforms become smoother and signals dwell less often on both power rails.

Fig. 6 shows relative histogram of $V_{mtj}$ with various low-pass filtering. The un-filtered signal shows high distribution at both power rails leaving low distribution in between (black). With increasing amount of filtering, distribution of $V_{mtj}$ becomes flatter (yellow), and even single-peaked (red).

As a result, since the input signal of p-block acts as a threshold to $V_{mtj}$ signal, the transfer curve of p-block, shown in fig. 7, changes from step-shape (black) without fileting to a more linear (yellow) and then sigmoidal (red) shape with more filtering. The sigmoidal transfer curve is more desirable for intended p-circuit operations.

### D. p-Block Response Time

To reveal the response time of the p-block, a step signal is fed to the p-block's input. Since the output is random, 100 output waveforms are captured and overlay plotted to observe the earliest-possible response of p-block at the output. As shown in fig. 8, the response time of p-block is well below 1 μs. This measurement is limited by the limited slew rate of the input step signal instead of the p-block circuit. Additionally, since the

MTJ oscillates randomly on the order of 1 MHz, limited number of waveform captures may slightly under-estimate the actual response time. Also note that the low-pass filtering on $V_{mtj}$ signal does not affect response time of p-block.

## III. DEMONSTRATION OF P-CIRCUIT IMPLEMENTING AN INVERTIBLE AND GATE

### A. Invertible AND Gate Functionality and Design

As shown in fig. 9(a), terminal A, B and C satisfy the relationship of C=AB. Compared to a standard AND gate, all three terminals of the invertible AND gate are both input and output capable. When no information is given to the gate A, B and C fluctuate seemingly randomly. But they follow the constraint of C=AB. When some information is given to some terminal(s), the state of that (these) terminal(s) become certain. Then other terminal(s) become either certain if only one state is legal according to the constraint or otherwise fluctuate among all legal states allowed by the constraint. The right of fig. 9(a) highlights the fact that the flow of information is bidirectional among all three terminals.

The three terminals of the invertible AND gate above are represented by three p-bits and implemented by three p-blocks. Schematics of the demonstration of this 'p-circuit' is shown in Fig. 9(b). Three p-blocks are built involving three MTJs. The outputs of p-blocks are coupled to their inputs via a resistor network. Values of resistors determine the intended functionality of the p-circuit, in this case being an invertible AND gate. Biasing resistors to both power rails are also parts of the resistor network. Information (clamp signals at 'clamp A, B, C') from outside of the p-circuit can enter via three clamp resistors, $R_{cA, B, C}$. If a clamp terminal is driven to $V_{dd}$ or $V_{ss}$, the corresponding p-bit is clamped to logic '0' or '1' respectively. If the clamp terminal is left floating, its p-bit is left free. Final states of the p-bits can be read at the p-block outputs.

Fig. 10 shows the waveform of all three p-block outputs when their inputs are disconnected from resistor network and to a ramp signal, 'test input'. This shows all three p-blocks working properly simultaneously. Note that the actual voltage of p-block outputs is between -1.65 V and 1.65 V. We consider voltage greater than 0 V as logic '1' and otherwise logic '0'.

### B. Invertible AND Gate Forward Operations

States of p-bits are read out by an oscilloscope at outputs of corresponding p-blocks. The combination of states of p-bit A, B and C are coded as [ABC] and represented as a 3-bit binary values ranging from '000' to '111' for all histogram results. And all states satisfy the AND constraint (legal) are plotted green while those violate the constraint (illegal) are plotted red.

Fig. 11 shows such relative histogram with A and B clamped to various certain states, as if information is flowing 'forward' through the AND gate. The results reflect the p-circuit's AND gate behavior as expected. Note that 'clamp=0(1)' means the clamp terminal is driven so that the corresponding p-bit is clamped to logic '0(1)'; 'clamp=n' means the terminal is left floating and the p-bit is left free.

Fig. 12 shows results when only one of A and B is clamped so that information flows 'forward' but is 'incomplete' on the standard 'input' side. When A is clamped to '0', B fluctuates between '0' and '1' and C always stay '0' being legal.

### C. Invertible AND Gate Backward Operations

For an invertible logic gate, information can also flow 'backwards' from the standard 'output' C side to the standard 'input' A and B side. As shown in Fig. 13, When C is clamped to '1', only [AB]=11 is legal, and as a result, all other combinations of [AB] are eliminated or suppressed. And when C is clamped to '0', three legal combinations of [AB] are present while the illegal combination '11' is eliminated.

### D. p-Circuit Stabilization Time

Following the case of C being clamped, fig. 14 shows waveforms of C, A and B when clamp signal at C is switched from $V_{ss}$ to $V_{dd}$ and C is transitioning from '1' to '0'. Before the transition when C is '1', [AB] are mostly '11'. After the transition of C to '0', A and B are more 'uncertain' but also is exclusive to each other and avoid the illegal '11' state. The effect of C on A and B is instant (on the order of few µs).

Fig. 15. shows the time evolution of [ABC] relative histogram with various sample time after C's transition. The histogram becomes correct and stable after 100 µs. Since the p-block's response time is much faster that 100 µs, this stabilization time is more attributed to the statistical accumulation process (uncertainty) of sampling A, B and C, considering limited rate of random change of p-bits.

### E. Invertible AND Gate Other Operations

One terminal from the standard 'input' side and C from the standard 'output' side can also be clamped. Fig. 16 shows the case where C and B are clamped while A is left free. A fluctuates among all legal states. When [BC]=01 which is already illegal, A goes to '1' to try to satisfy the AND constraint. The mirrored cases where B is left free yield similar results.

### F. Invertible AND Gate Free-run Operation

When all three p-bit are left free, they fluctuate among all legal states under the AND constraint. As shown in fig. 17, all green legal states are more likely than the illegal states by adequately large margins.

## IV. CONCLUSIONS

The authors have experimentally demonstrated the MTJ-based PSL. The basic building block, p-block, and then a p-circuit implementing an invertible AND logic gate, are designed and demonstrated. Operation scheme of thermally stable MTJ involving bias voltage and field is adopted in this demonstration to compensate device-device variations.

ACKNOWLEDGMENT

This work is supported in part by E2CDA CAPSL/nCORE program and DARPA FRANC program.

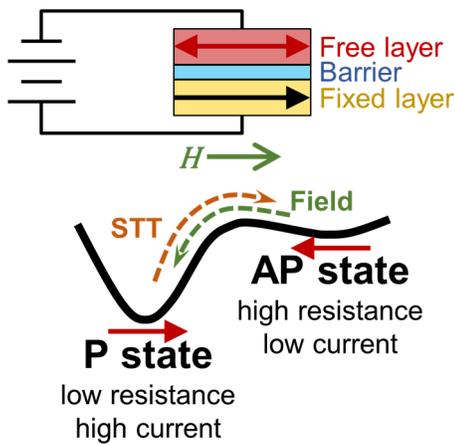

Fig. 1. Operation mechanism of random oscillation of thermally stable MTJ.

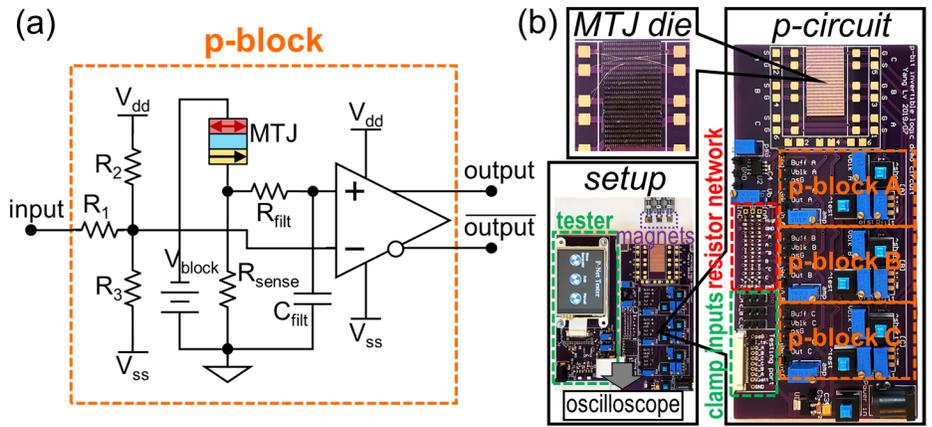

Fig. 2. (a) PSL p-bit building block (p-block) simplified circuit diagram. (b) Photo experimental setup, which consists of the p-circuit main board with an MTJ die mounted and wire-bonded, a tester board and magnets as bias field source. The p-circuit mainly consists of three p-blocks and a resistor network.

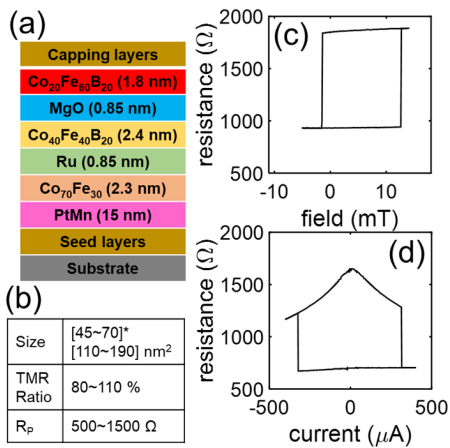

Fig. 3. MTJ (a) stack structure (b) key parameters (c) resistance vs. external field (d) resistance vs. current.

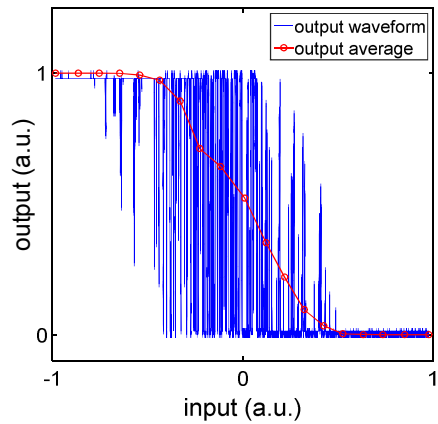

Fig. 4. p-block output waveform (blue) and output average (red) vs. input.

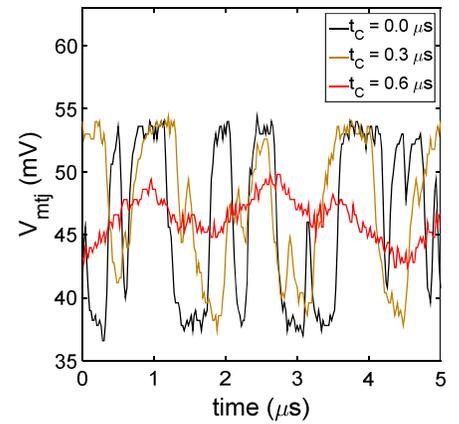

Fig. 5. MTJ current sensed by $R_{sense}$, $V_{mtj}$, waveform with various filtering strength (time constant, $t_C$).

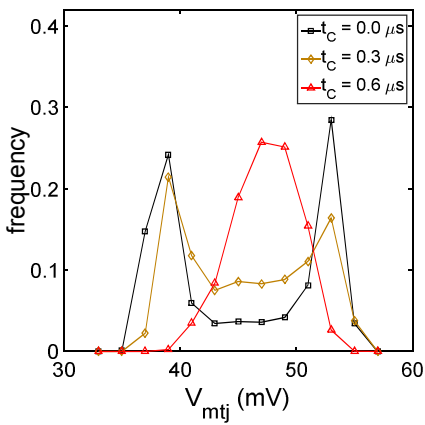

Fig. 6. MTJ current sensed by $R_{sense}$, $V_{mtj}$, relative histogram with various filtering strength (time constant, $t_C$).

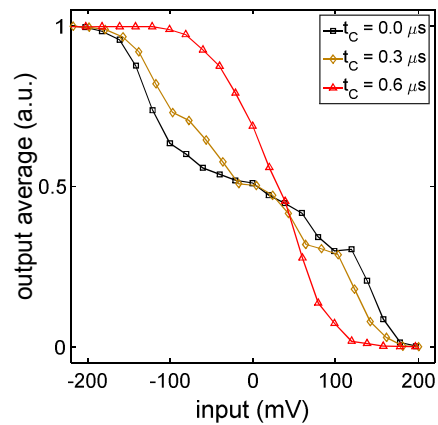

Fig. 7. p-block output average vs. input with various filtering strength (time constant, $t_C$).

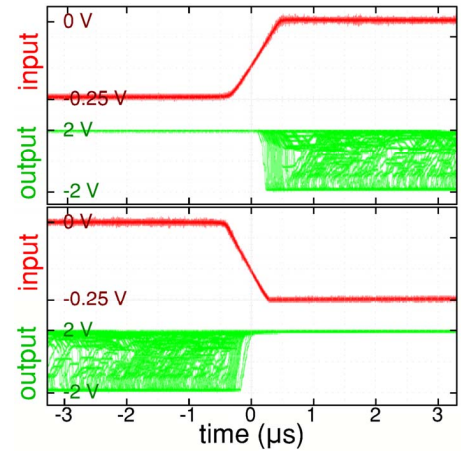

Fig. 8. p-block output step response waveform (100 captured waveforms overlay plotted).

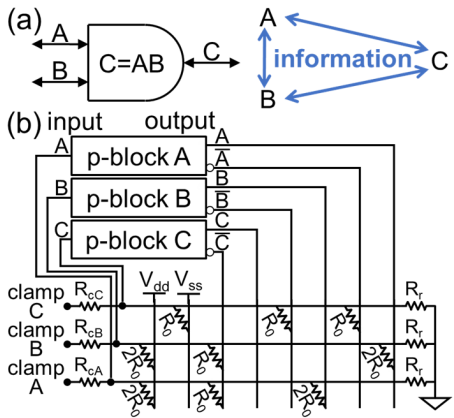

Fig. 9. (a) Illustration of invertible AND gate functionality and flow of information (b) simplified circuit diagram of p-circuit implementation of an invertible AND gate.

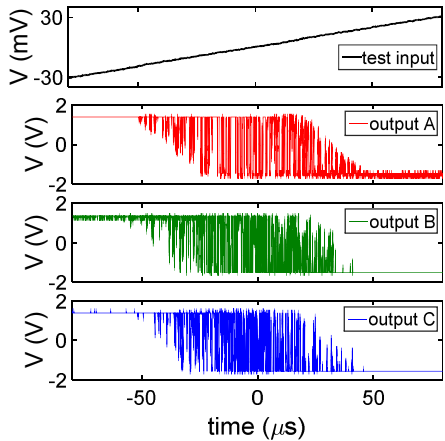

Fig. 10. Waveforms of a ramp input test signal and output A, B and C of all three p-bit blocks.

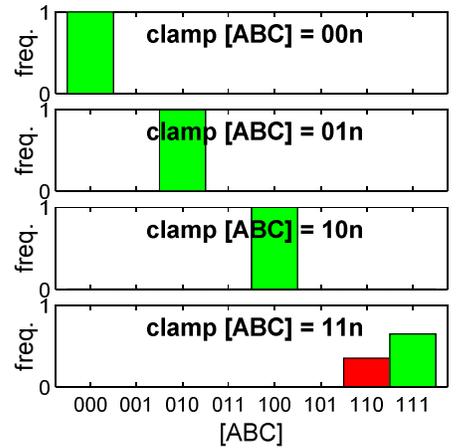

Fig. 11. Relative histogram of [ABC] with A and B clamped showing complete forward operations.

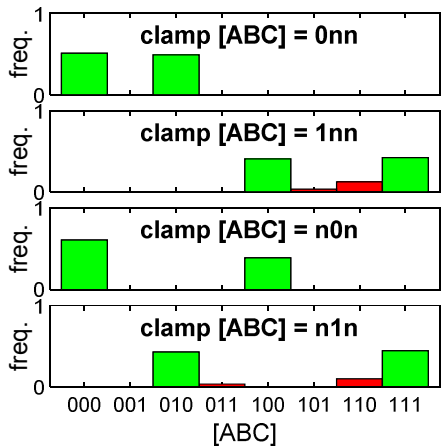

Fig. 12. Relative histogram of [ABC] with only A or B clamped showing incomplete forward operations.

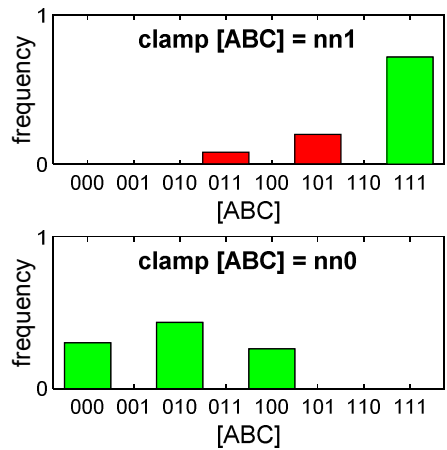

Fig. 13. Relative histogram of [ABC] with C clamped showing backward operations.

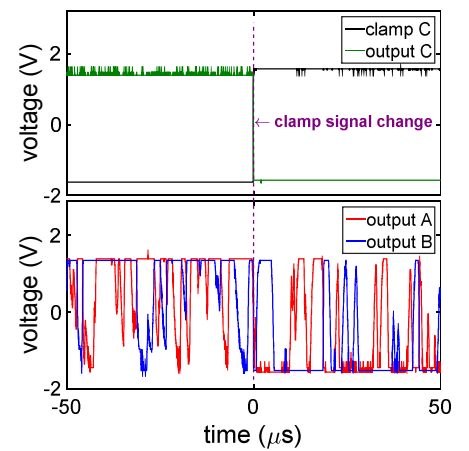

Fig. 14. Waveform of clamp C, and output A, B and C when C changes from '1' to '0'.

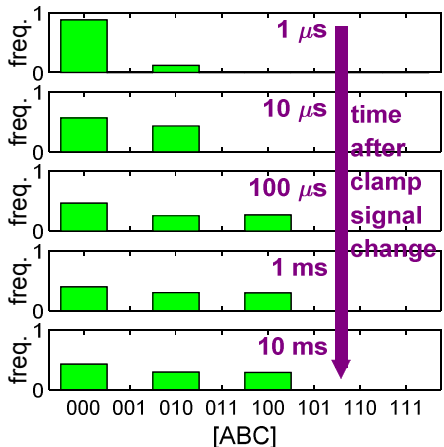

Fig. 15. Time evolution of [ABC] relative histogram showing [ABC] stabilization process after C changes from '1' to '0'.

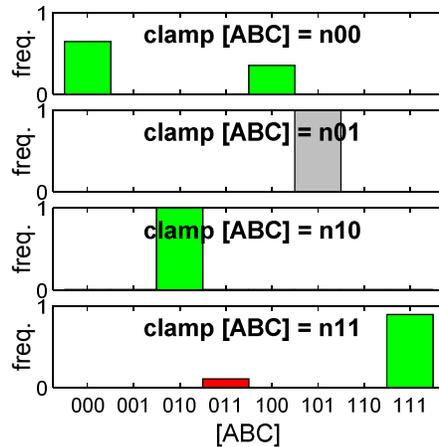

Fig. 16. Relative histogram of [ABC] with B and C clamped showing operations with information input from both sides.

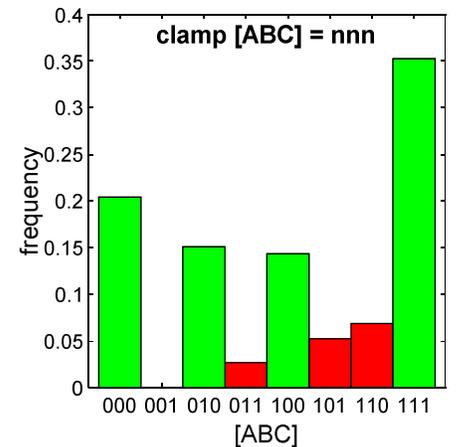

Fig. 17. Relative histogram of [ABC] with A, B and C left unclamped showing free-run operation.